\documentstyle[12pt]{article}
\catcode`\@=11

\@addtoreset{equation}{section}
\def\theequation{\arabic{equation}}

\def\@normalsize{\@setsize\normalsize{15pt}\xiipt\@xiipt
\abovedisplayskip 14pt plus3pt minus3pt%
\belowdisplayskip \abovedisplayskip
\abovedisplayshortskip  \z@ plus3pt%
\belowdisplayshortskip  7pt plus3.5pt minus0pt}
\def\small{\@setsize\small{13.6pt}\xipt\@xipt
\abovedisplayskip 13pt plus3pt minus3pt%
\belowdisplayskip \abovedisplayskip
\abovedisplayshortskip  \z@ plus3pt%
\belowdisplayshortskip  7pt plus3.5pt minus0pt
\def\@listi{\parsep 4.5pt plus 2pt minus 1pt
            \itemsep \parsep
            \topsep 9pt plus 3pt minus 3pt}}

\def\underline#1{\relax\ifmmode\@@underline#1\else
        $\@@underline{\hbox{#1}}$\relax\fi}
\@twosidetrue
\relax

\catcode`@=12

\catcode`\@=11

\def\section{\@startsection{section}{1}{\z@}{3.5ex plus 1ex minus
   .2ex}{2.3ex plus .2ex}{\large\bf}}

\def\ps@headings{\def\@oddfoot{}\def\@evenfoot{}
\def\@oddhead{\hbox{}\hfill
        \makebox[.5\textwidth]{\raggedright\ignorespaces --\thepage{}--
        \hfill }}
\def\@evenhead{\@oddhead}
\def\subsectionmark##1{\markboth{##1}{}}
}

\ps@headings

\catcode`\@=12

\relax

\def\figcap{\section*{Figure Captions\markboth
        {FIGURECAPTIONS}{FIGURECAPTIONS}}\list
        {Fig. \arabic{enumi}:\hfill}{\settowidth\labelwidth{Fig. 999:}
        \leftmargin\labelwidth
        \advance\leftmargin\labelsep\usecounter{enumi}}}
 \relax
\def\tablecap{\section*{Table Captions\markboth
        {TABLECAPTIONS}{TABLECAPTIONS}}\list
        {Table \arabic{enumi}:\hfill}{\settowidth\labelwidth{Table 999:}
        \leftmargin\labelwidth
        \advance\leftmargin\labelsep\usecounter{enumi}}}
 \relax
\def\reflist{\section*{References\markboth
        {REFLIST}{REFLIST}}\list
        {[\arabic{enumi}]\hfill}{\settowidth\labelwidth{[999]}
        \leftmargin\labelwidth
        \advance\leftmargin\labelsep\usecounter{enumi}}}
 \relax

\catcode`\@=11

\def\marginnote#1{}
\newcount\hour
\newcount\minute
\newtoks\amorpm
\hour=\time\divide\hour by60
\minute=\time{\multiply\hour by60 \global\advance\minute by-
\hour}
\edef\standardtime{{\ifnum\hour<12 \global\amorpm={am}%
    \else\global\amorpm={pm}\advance\hour by-12 \fi
    \ifnum\hour=0 \hour=12 \fi
    \number\hour:\ifnum\minute<100\fi\number\minute\the\amorpm}}
\edef\militarytime{\number\hour:\ifnum\minute<100\fi\number\minute}
\def\draftlabel#1{{\@bsphack\if@filesw {\let\thepage\relax
  \xdef\@gtempa{\write\@auxout{\string
    \newlabel{#1}{{\@currentlabel}{\thepage}}}}}\@gtempa
    \if@nobreak \ifvmode\nobreak\fi\fi\fi\@esphack}
     \gdef\@eqnlabel{#1}}
\def\@eqnlabel{}
\def\@vacuum{}
\def\draftmarginnote#1{\marginpar{\raggedright\scriptsize\tt#1}}
\def\draft{\oddsidemargin -.5truein
        \def\@oddfoot{\sl preliminary draft \hfil
        \rm\thepage\hfil\sl\today\quad\militarytime}
        \let\@evenfoot\@oddfoot \overfullrule 3pt
        \let\label=\draftlabel
        \let\marginnote=\draftmarginnote
   
\def\@eqnnum{(\theequation)\rlap{\kern\marginparsep\tt\@eqnlabel}%
\global\let\@eqnlabel\@vacuum}  }
\def\preprint{\twocolumn\sloppy\flushbottom\parindent 1em
        \leftmargini 2em\leftmarginv .5em\leftmarginvi .5em
        \oddsidemargin -.5in    \evensidemargin -.5in
        \columnsep 15mm \footheight 0pt
        \textwidth 250mmin      \topmargin  -.4in
        \headheight 12pt \topskip .4in
        \textheight 175mm
        \footskip 0pt
        
\def\@oddhead{\thepage\hfil\addtocounter{page}{1}\thepage}
        \let\@evenhead\@oddhead \def\@oddfoot{} \def\@evenfoot{} 
}
\def\titlepage{\@restonecolfalse\if@twocolumn\@restonecoltrue\onecolumn
     \else \newpage \fi \thispagestyle{empty}\c@page\z@
        \def\thefootnote{\fnsymbol{footnote}} }
\def\endtitlepage{\if@restonecol\twocolumn \else  \fi
        \def\thefootnote{\arabic{footnote}}
        \setcounter{footnote}{0}}  
\catcode`@=12
\relax

\def\ps@headings{\def\@oddfoot{}\def\@evenfoot{}
\def\@oddhead{\hbox{}\hfill
        \makebox[.5\textwidth]{\raggedright\ignorespaces --\thepage{}--
        \hfill }}
\def\@evenhead{\@oddhead}
\def\subsectionmark##1{\markboth{##1}{}}
}

\ps@headings

\relax

\def\firstpage#1#2#3#4#5#6{
\begin{document}
\def\beq{\begin{equation}} 
\def\eeq{\end{equation}} 
\def\bea{\begin{eqnarray}} 
\def\eea{\end{eqnarray}} 
\def\bq{\begin{quote}} 
\def\eq{\end{quote}}
\def\ra{\rightarrow} 
\def\lra{\leftrightarrow} 
\def\ups{\upsilon}
\def\bq{\begin{quote}} 
\def\eq{\end{quote}}
\def\ra{\rightarrow} 
\def\un{\underline}
\def\ov{\overline}
\newcommand{\cm}{Commun.\ Math.\ Phys.~}
\newcommand{\prl}{Phys.\ Rev.\ Lett.~}
\newcommand{\pr}{Phys.\ Rev.\ D~}
\newcommand{\pl}{Phys.\ Lett.\ B~}
\newcommand{\ibar}{\bar{\imath}}
\newcommand{\jbar}{\bar{\jmath}}
\newcommand{\np}{Nucl.\ Phys.\ B~}
\newcommand{\F}{{\cal F}}
\renewcommand{\L}{{\cal L}}
\newcommand{\A}{{\cal A}}
\def\154{\frac{15}{4}}
\def\153{\frac{15}{3}}
\def\32{\frac{3}{2}}
\def\254{\frac{25}{4}}
\begin{titlepage}
\nopagebreak
\title{\begin{flushright}
        \vspace*{-1.8in}
        {\normalsize CERN-TH/97-311}\\[-9mm]
        {\normalsize IOA--TH.97-15}\\[-9mm]
        {\normalsize UUTP-22/97}\\[-9mm]
        {\normalsize hep-th/9711044}\\[4mm]
\end{flushright}
\vfill
{#3}}
\author{\large #4 \\[1.0cm] #5}
\maketitle
\vskip -7mm     
\nopagebreak 
\begin{abstract}
{\noindent #6}
\end{abstract}
\vfill
\begin{flushleft}
\rule{16.1cm}{0.2mm}\\[-3mm]
$^{\dagger}${\small Supported by the European Community under Human 
Capital and Mobility Grant No ERBCHBICT960773}\\
CERN-TH/97-331\\
October 1997
\end{flushleft}
\thispagestyle{empty}
\end{titlepage}}
\def\simlt{\stackrel{<}{{}_\sim}}
\def\simgt{\stackrel{>}{{}_\sim}}
\date{}
\firstpage{3118}{IC/95/34}
{\large\bf On The Instanton Solutions Of 
The Self-Dual Membrane\\ In Various Dimensions} 
{E.G. Floratos$^{\,a,b}$, G.K. Leontaris$^{\,c,d}$, A.P. Polychronakos$^{\,c,e}$
and R. Tzani$^{\,c\dagger}$}
{\normalsize\sl
$^a$ NRCS Demokritos, Athens, Greece\\[-3mm]
\normalsize\sl
$^b$ Physics Department, University of Iraklion,
Crete, Greece.\\[-3mm]
\normalsize\sl
$^c$Theoretical Physics Division, Ioannina University,
GR-45110 Ioannina, Greece\\[-3mm]
\normalsize\sl
$^d$CERN, Theory Division, 
1211 Geneva 23, Switzerland\\[-3mm]
\normalsize\sl
$^e$Theoretical Physics Department, Uppsala University,
S-751 08 Uppsala, Sweden.}
{We present some methods of determining explicit solutions for self-dual
supermembranes in $4+1$ and $8+1$ dimensions with spherical or 
toroidal topology. For configurations of axial symmetry, 
the continuous $SU(\infty)$ Toda equation
turns out to play a central role, and a specific method of determining all the
periodic solutions are suggested. A number of examples are studied in detail. 
}

\newpage
Nowadays, a revived interest in membrane theory~\cite{1} 
has been spurred by the fact that M-theory, which is considered as the 
leading candidate theory for explaining the non-perturbative net of
string dualities,  contains membranes and 
their dual five-branes in eleven dimensions~\cite{2}.
 The main activity in recent literature has been the classification of
the BPS spectra of various string compactifications which pressumably
M-theory owes to organize in a compact and intuitive way.  
Among the BPS states, there is an important class made up of the Euclidean
solitons  (instantons). This sector plays a role in the understanding 
of the non-perturbative vacuum structure of string compactifications. 

Some years ago, we  introduced, at the level of the bosonic membranes,
a specific self-duality, which in modern language is nothing but S-duality 
for Euclidean instantons~\cite{fl1}. The self-dual membranes  solve
$SU(N)$ Nahm's equations for a specific  $N\ra \infty $ limit where 
$SU(N)$ becomes the area-preserving diffeomorphism group on the 
surface of the membrane, a symmetry that exists in the light-cone
quantization of the membranes. 

Recently, extensions of the self-duality
of membranes in 7,8,9 dimensions have been introduced~\cite{cfz,fl}.
In the present work, we develop new methods for solving the self-duality equations
in three and seven dimensions~\cite{fair}.  In the case of toroidal compactifications,
the role of string excitations of self-dual membranes becomes visible
and we exhibit explicit examples where analytic solutions are found.
In three, and also in seven dimensions, and for the case of
cylindrical symmetry, the self-duality equations reduce to continuous
Toda equations which have been studied in order to determine self-dual
Euclidean solutions of Einstein equations~\cite{Toda}.
In the present work, we provide a first-order non-linear system,
the axially-symmetric three-dimensional self-duality equations,
which at the same time provide a Lax pair of the axially symmetric
Toda equation. Inverting this non-linear system, we find a completely
integrable linear system, which we explicitly solve and, thus, we 
present a method to determine all the solutions of the axially
symmetric Toda equations~\cite{RW,BS}.

We start our analysis by reviewing the salient features of the theory.
In ref.~\cite{FIT} it was pointed out that in the large-$N$ limit,
$SU(N)$ YM theories have, at the classical level, a simple
geometrical structure with the $SU(N)$ matrix potentials $A_{\mu}(X)$ replaced by
c-number functions of two additional cordinates $\theta, \phi$ of an internal
sphere $S^2$ at every space time point,  while the $SU(N)$ symmetry is replaced 
by the infinite-dimensional algebra of area-preserving diffeomorphisms of the
sphere $S^2$ called $SDiff (S^2)$. 
The $SU(N)$ fields are Hermitian $N\times N$ matrices which in the large
$N$ limit are written in terms of the spherical harmonics on $S^2$,
while commutators are replaced by the Poisson brackets on $S^2$:
\begin{eqnarray}
[A_{\mu},A_{\nu}]&\ra&\{A_{\mu},A_{\nu}\}\nonumber\\
&=&\frac{\partial A_{\mu}}{\partial\sigma_1}
\frac{\partial A_{\nu}}{\partial\sigma_2}
-\frac{\partial A_{\nu}}{\partial\sigma_1}
\frac{\partial A_{\mu}}{\partial\sigma_2}. \label{6}
\end{eqnarray}
In three dimensions the self-duality relation is defined
by  the equation\footnote{The anti-self dual case $E_i=-B_i$ can be
treated similarly.}
\begin{equation}
E_i= B_i,
\end{equation}
where
$E _ i$ and $ B _ i$ are the electric and the magnetic $SU(\infty )$
colour fields. 
Since
\begin{equation}
E_ i = \frac{\partial A _ i} {\partial t}, \quad
i=1,2,3 \label{12}
\end{equation}
and
\begin{equation}
B_i=\frac{1}{2}\varepsilon_{ijk}\{A_j,A_k\}, \label{13b}
\end{equation}
where $\varepsilon_{ijk}$ is the antisymmetric tensor in three dimensions,
one obtains the following  equations
\begin{equation}
\dot{A}_i=\frac{1}{2}\varepsilon_{ijk}\{A_j,A_k\},
\quad i,j,k=1,2,3. \label{16}
\end{equation}

These equations solve the Gauss constraints and the second-order Euclidean equations
of motion for the bosonic part of the supermembrane (fermionic DOF set to zero)
in the light-cone gauge~\cite{rev}.
In what follows we discuss methods of solution of the above three-dimensional system.

It has been suggested in ref.~\cite{fl1} that one can use quaternions to transform
the above equations into a matrix differential one. 
We define the matrix 
\beq
A = A_i \sigma_i, \, i=1,2,3,
\label{AM}
\eeq
where $\sigma_i$ are the standard Pauli matrices. The matrix function
$A$ satisfies the equation
\beq
\dot{A}= -\frac{\imath}2\{A,A\}.
\label{ME}
\eeq
In the case of the sphere, which has been analysed in ref.~\cite{fl1},
the Darboux coordinates are $\xi_1=\cos\theta$, $\xi_2=\phi$. 
The infinite-dimensional group $SDiff(S^2)$ has $SO(3)$ 
 as the only finite-dimensional subgroup which is generated 
by the three functions $e_1= \cos\phi \sin\theta$
$e_2=\sin\phi \sin\theta$, $e_3=\cos\theta$:
\beq
\{e_i,e_j\} = - \varepsilon_{ijk} e_k
\label{PBe}
\eeq
Looking for factorized $SO(3)$-symmetric solutions, we set $A= T_i(t)e_i$, which
implies 
\beq
\dot{T}_i = \frac{\imath}4 \varepsilon_{ijk}[T_j,T_k],
\label{nahm}
\eeq
that is, the Nahm equation for an $SU(2)$ monopole of magnetic charge
$k=2$~\cite{Nahm}.
We thus obtain for each choice of solution of the Nahm equations for
magnetic charge $k=2$ (eight-dimensional moduli space) a solution of the
self-duality equations.

This system of equations is known to be integrable, and particular solutions
for specific boundary conditions at $t=0$, $t=2$ (simple poles with
$SU(2)$ matrices as residues) can be expressed in terms of elliptic
functions~\cite{sut}.  In ref.~\cite{fl1}, zero total angular momentum 
(axially symmetric) solutions
of the system (\ref{16}) have been explicitly determined in terms of
the functions $e_i$ and the solutions of the $SU(2)$ Toda equation.

In the following we will show that the requirement of axial symmetry on
 the above system leads to a first-order system for two functions, 
which plays the role of the Lax pair for the continuous axially symmetric
Toda equation. Indeed, the ansatz 
\bea
A_1  = R(\sigma_1,t) \cos \sigma_2 ,&
A_2 = R(\sigma_1,t) \sin \sigma_2 ,&
A_3 = z(\sigma_1,t)
\eea
leads to the system,
\bea
\dot{z}&=& R R'\label{Zeq}\\
\dot{R}&=& - R z'\label{Req}
\eea
where the prime now is used to declare differentiation with respect to $\sigma_1$
(i.e. $ \frac {\partial}{\partial \sigma _ 1} $)\footnote{
We observe that, if we replace $\sigma_2$ by $n \sigma_2$, $n$ integer, then
this implies that $t\ra n t$ in the original solution.}.
 Combining equations (\ref{Zeq}) and (\ref{Req}) we obtain the 
axially symmetric continuous Toda  equation
\beq
\frac{d^2\Psi}{d t^2}+ \frac{d^2{e^{\Psi}}}{d \sigma_1^2}=0,
\label{,.}
\eeq
where  $R^2 = e^{\Psi}$. Solutions of this equation have been
discussed in the literature in connection with the self-dual 4d Einstein
metrics with rotational and axial Killing vectors~\cite{Toda,RW}.
Here though, we note that $\sigma_1$ runs in a compact interval
($0, 2\pi$) for torus and ($-1,1$) for the sphere.

At this point, we want to provide a specific example of a solution
with separation of variables of the Toda equation, in
the case of spherical topology ($\sigma_1 =\cos\theta$, $\sigma_2=\phi$).
Separation of variables $R(\sigma_1,t)= R_1(\sigma_1)R_2(t)$
corresponds to $\Psi (\sigma_1,t)= \Theta (\sigma_1)+ T(t)$ which
leads to
\bea
\frac{d^2T}{d t^2} - k  e^T =0\label{T1}\\
\frac{d^2e^{\Theta}}{d \sigma_1^2} +k   =0\label{The1}
\eea
Multiplying (\ref{T1}) by $\dot{T}$ we obtain
\beq
\frac{dT}{dt} = \sqrt{2 k}\left( e^T +\frac{\nu}{k}\right)^{1/2}
\label{T2},
\eeq
where $\nu$ is a new constant. Equation (\ref{T2}) is easily solved, making
use of the transformation $e^T = \frac{\nu}{k}\phi^{-2}$,  
the final result being
\bea
R(\theta,t)&=&  \kappa\frac{\sin\theta}{\sinh\left[\kappa
(t_0-t)\right]}\\
z(\theta,t)&=&  \kappa \coth\left[\kappa
(t_0-t)\right]\cos\theta,
\eea
where $\kappa$ is a new constant. Interestingly, this solution
coincides with that of ~\cite{fl1} representing axisymmetric
ellipsoids,
which was derived from the $SU(2)$ Toda equation (with respect
to the time $t$).

We now exhibit a variation of the method of ref.~\cite{ward}     
where by inversion of the non-linear system (\ref{Req}) 
we construct a linear one and we determine all solutions.
Indeed, by going from the pair of variables ($R,z$) to ($S,T$),
which we take to be the inverse mapping $(\sigma_1,t)\ra (R,z)$,
we find 
\bea
\frac{\partial S }{\partial u} -  \frac{\partial T}{\partial v }& =& 0\\
\frac{\partial S}{\partial v} + u \frac{\partial T }{\partial u}&=&0\label{inv}, 
\eea
where $u=R^2$ and $v=2 z$.
This system is linear and we can easily separate the variables $u$ and $v$,
$S = S_1(u) S_2(v),\,\, T = T_1(u) T_2(v)$.
We introduce two constants of separation, 
\bea
\begin{array}{cc}
{{\partial S_1 }/{\partial u}=\lambda {T_1}},&
-u {\partial T_1}/{\partial u}=\mu S_1 \\
{{\partial T_2}/{\partial v}=\lambda {S_2}},&  
{\partial S_2 }/{\partial v}=\mu T_2.
\end{array}
\label{sep}
\eea
We see that $S_2$ and $T_2$ are trigonometric (hyperbolic) functions of $v$
depending on whether the sign of the product $\lambda\cdot\mu$ is minus (plus).
Also, from the first order equations of $S_1$ and $T_1$, assuming analyticity
around $u=0$, we obtain unique solutions $T_1\propto J_0(k_0 R)$ and 
$S_1\propto R J_1(k_0 R)$, and $k_0 = \sqrt{\lambda\cdot\mu}$. 
By appropriate linear combinations
of the solutions of $S_1,T_1$ and $S_2,T_2$, we can determine  
functions $S$ and $T$
which, by inversion, give functions $R,z$, periodic in $\sigma_1$.
As a demonstration, consider the solution
\bea
S = \imath A \cos (k_0 z) R J_1 (\imath k_0 R) + k_1 z \\
T = A \sin (k_0 z) J_0 (\imath k_0 R) - k_1 \ln R 
\eea
where $A,k_1,k_0$ are real constants. If space is compactified in 
the $z$-direction with length $L$, 
and we want $\sigma_1$ to range from 0 to $2\pi$, we
choose $k_1 =2\pi/L$ and $k_0 =nk_1 $ for some integer $n$. 
The above then represents a membrane with $n$ branches extending 
to $R=\infty$,  which, at some critical
time, collides with itself and separates into a finite piece with
toroidal topology, exhibiting $n$ ripples within the period $L$, and $n$
infinite pieces that fly away. We leave the question of explicit 
constructions for a future work. We should note, though, that
the linearization method of ref.~\cite{ward} should be examined in
more detail in order to construct other interesting examples.

We finally discuss in three dimensions two sorts of toroidal 
compactifications where by double compactification we derive 
string self-dual solutions. First, when the three dimensional space
topology is $R^2\times S^1$, we doubly compactify the membrane
~\cite{1}. We choose as an example  $A_3=n \sigma_2$ and $A_{1,2}= A_{1,2}(\sigma_1,t)$.
Then it is straightforward to see that $A_{1}+\imath A_2$ must be
an analytic function of $\sigma_1-\imath n  t$, where $n$ is 
the winding number. These are world-sheet
string instantons.

 The second compactification is on the three-dimensional
torus $T^3$,
 where windings for various embeddings of toroidal membrane 
lead to string excitations with non-zero center-of-mass momentum.
We discuss this case below, where more general seven-dimensional embeddings
are studied. 
We now extend  our discussion in seven dimensions, where the fully antisymmetric
 symbol of three dimensions 
 $\varepsilon_{ijk}$ in  eqs.(\ref{16}) is replaced by the
corresponding octonionic structure constants $\Psi_{ijk}$~\cite{cfz,fl}:
\begin{equation}
\dot{X}_i = \frac{1}{2} \Psi_{ijk}\{X_j,X_k\},
\label{osce}
\end{equation}
where the indices run from 1 to 7 while $\Psi_{ijk}$ is completely antisymmetric 
 and has the value 1 for the following combinations of indices:
\beq
 \Psi_{ijk}=\left\{\begin{array}{ccccccc}1&2&4&3&6&5&7\\
                             2&4&3&6&5&7&1\\
                             3&6&5&7&1&2&4
 \end{array}\right.
\label{2.1}
\eeq
The second-order Euclidean equations and the
 Gauss law results automatically by making use of the $\Psi_{ijk}$ 
cyclic symmetry
$\{\dot{X}_i,X_i\}= 0$.
In ref.~\cite{fl}, one class of three-dimensional solutions which are embedded in the 
seven-dimensional system was found according to the identifications
\beq
X_3\ra A_3, \;\;\; X_{\pm}\ra A_{\pm}/\sqrt{3}\label{7to3}
\eeq
where the seven coordinates $X_{i},(i=1,2,...,7)$ are grouped  
in terms of the complex
coordinates $X_{\pm} =X_1\pm \imath X_2,$,  $Y_{\pm} =X_4\pm \imath X_5,$
and  $Z_{\pm} =X_6\pm \imath X_7$ and we have made the ansatz that
${X}_+=Z_+ = \imath Y_-$, while $A_{\pm,3}$ is the three-dimensional solution.
The  seven-dimensional solution is essentially the three-dimensional one rotated
by an orthogonal transformation in 7-space. 
Therefore, any  three-dimensional self-dual solution automatically generates
a corresponding 7 dimensional one.

The generalization to the string-like solution of the self-duality
equation (\ref{osce}) in 7 dimensions is straightforward. 
We assume the form
\beq
X_i(\sigma_{1,2},t) = A_i \sigma_1 +B_i \sigma_2 +P_i t+
f_i(\sigma_1,\sigma_2,t)
\eeq
with $i=1,...,7$, and $f$ being a periodic function of $\sigma_{1,2}$
and $A,B$ integer vectors.
Then we obtain
\bea
P_i & =& \Psi_{ijk} A_j B_k\label{Pi}\\
\dot{f}_i & = & \Psi_{ijk} \left( A_j \frac{\partial}{\partial\sigma_2}-B_j
 \frac{\partial}{\partial\sigma_1}\right) f_k +\frac{1}{2}\Psi_{ijk}\{f_j,f_k\}
\label{Pi1}
\eea
Since $f$ is a periodic function with respect to $\sigma_{1,2}$, 
we can expand it in terms of an infinite number of strings, depending on the
coordinate $\sigma_1$: 
\beq
f_i(\sigma_1,\sigma_2,t) = \sum_nX_i^n(\sigma_1,t) e^{in\sigma_2}.
\eeq
Then, from the self-duality equations (\ref{Pi},\ref{Pi1})
 we find that the winding number of the
membrane is related to the center-of-mass momentum, which is transverse to
the compactification directions $A$ and $B$. Also, the infinite   
number of strings are coupled through the following equations
\beq
\dot{X}_i^n(\sigma_1,t) = \Psi_{ijk} \left(A_j n - B_j 
\frac{\partial}{\partial\sigma_1}\right)
X_k^n +\frac{ \imath}{ 2}\Psi_{ijk}
\sum_{n_1+n_2=n}\left(n_2 
\frac{\partial X_j^{n_1}}{\partial\sigma_1}X_k^{n_2}-n_1X_j^{n_1}
 \frac{\partial X_k^{n_2}}{\partial\sigma_1}\right)
\eeq
The string-like solution corresponds to the particular case
$ {\partial f_i}/{\partial \sigma_2}=0$, where we obtain
\beq
X_i^0 = X_i(\sigma_1,t)\ra \dot{X}_i = \Psi_{ijk}B_k 
\frac{\partial X_j}{\partial\sigma_1}.
\eeq
This equation is formally solved in vector form by
\beq
X (\sigma_1, t) = e^{t M \frac{\partial }{\partial\sigma_1}}
X (\sigma_1, 0)
\label{form}
\eeq
where we defined the $7 \times 7$ matrix
$M_{ij} = \Psi_{ijk}B_k$. Explicit solutions are found by
expanding $X_i$ in terms of the eigenvectors of $M$.
In fact, since $M$ is real and antisymmetric, the real 7-dimensional
vector space decomposes into three orthogonal two-dimensional subspaces, 
each corresponding to a pair of 
imaginary eigenvalues $\pm \imath\lambda$, and a one-dimensional
subspace, in the direction of $B_i$, corresponding to the zero eigenvalue.
Since, in addition, $(M^2)_{ij} = - B^2 \delta_{ij} + B_i B_j$ (as
can be checked), we see that the imaginary eigenvalue pairs are all
$\pm \imath |B|$. Therefore the problem decomposes into three 3-dimensional
problems (one for each subspace) of the kind we solved before. The
general solution is then
\bea
&X_1^{(n)} +\imath  X_2^{(n)} = F_n ( \sigma_1 -\imath B t ) ~,~~~n=1,2,3\\
&X^{(0)} = |B| t
\eea
where $(X_1^{(n)} , X_2^{(n)} )$ are the projections of the membrane
coordinates on the $n$-th two-dimensional eigenspace and $X^{(0)} =
X_i B_i / |B|$ is the projection on $B_i$. As an example, if we choose
$B_i$ in the third direction, $B_i = B \delta_{i3}$, we have
\bea 
&X_1 + i X_2 = F_1 ( \sigma_1 -\imath B t )\\
&X_5 + i X_4 = F_2 ( \sigma_1 -\imath B t )\\
&X_6 + i X_7 = F_3 ( \sigma_1 -\imath B t )\\
&X_3 = B t.
\eea

Considering, now, the case when at $t=0$ we have
a proper (not string-like) membrane configuration,
with its periodic part dependent on both variables, we write
\beq
X_i = f_{i}^{cl}+ f_i,
\eeq
where $f_i^{cl} = A_i \sigma_1 +B_i \sigma_2$. The equation of the general 
case (\ref{osce}) can be written in a symbolic form, by defining the matrix
differential operator 
\beq
L_f^{ik} = \Psi_{ijk} \left(\frac{\partial f_j}{\partial\sigma_1}
\frac{\partial}{\partial\sigma_2}-
\frac{\partial f_j}{\partial\sigma_2}
\frac{\partial}{\partial\sigma_1}\right)
\label{oper}
\eeq
as a vector equation 
\beq
\dot{f} = (L_{f^{cl}} +\frac 12 L_f) f.
\eeq
It is possible to solve this non-linear matrix differential system by 
iteration of the solution of its linear part,
\beq
\dot{g}= L_{f^{cl}} g .
\label{homg}
\eeq
The above differential system can be written as a matrix integral equation 
as follows  
\beq
f= g + \frac 12e^{t L_{f^{cl}}}
\int^t  e^{-t' L_{f^{cl}}} L_f f dt'\label{inte}
\eeq
It is easy to show that the infinite iteration of the solution $g$ solves 
the non-linear differential system and, moreover, when the initial configuration 
is a string, the second part of the integral equation is zero and the problem is
reduced to the homogeneous case.  The general solution of the homogeneous
system (\ref{homg}) is
\beq
g(t) = e^{t L_{f^{cl}}} g(t=0),
\eeq
where $f_i(t=0)= g_i(t=0)$.

At this point, we would like to note that in ref~\cite{fl}
for  the case of
zero-winding we have been able to separate the time and the
parameter dependence of the coordinates of the octonionic
self-dual membrane. The time equations are generalizations
of Nahm matrix equations (\ref{nahm}), where in the place of the three
$SU(2)$-$T_i$ matrices, a pair $T_i, S_i$ appears.
A generalization of the Euler Top equations using octonions
has been proposed in ref~\cite{fair}, where  it was shown that this
system is an integrable one and the explicit set of seven
conservation laws including their algebraic relation has
been provided. This system of equations is a specific case of
the generalized Nahm equations when $T_i, S_i$ are proportional
to the Pauli matrices. Thus, for every solution of the generalized
Euler Top system,  one can obtain the corresponding self-dual membrane.

We close our short analysis by pointing out the existence 
of a different kind of self-duality equations which satisfy also
the second-order Euclidean equations which has been introduced for
self-dual Yang-Mills fields in ref.~\cite{iva}.\footnote{ We
thank T. Ivanova for bringing  her work with
A. Popov to our attention .} This system of equations could be generalized to 
membranes embedded in dimensions $D=dim(G)$ where $G$ is
any Lie algebra. This system of self-duality equations is an
integrable one (as it was  pointed out to us by T. Ivanova)
but the geometrical significance for the dynamics of the self-dual membrane
is not obvious to us. On the other hand, it is interesting to see
what type of world-volume membrane instantons are obtained  by
this method.

We would like to conclude with few remarks. A systematic approach
for the solutions of the seven-dimensional equations has been 
proposed in the case of toroidal compactifications which turn out
to provide  world volume membrane instantons which play an important role
in the understanding of the vacuum structure of supermembrane theory.
The question of the surviving supersymmetries for various classes
of solutions is an important problem for the determinations of
the BPS states of the supermembrane. The richness of the self-duality
equations concerning string excitations suggest that probably it is
the right framework of examining the non-perturbative  unification
of string interactions. This goes along the lines of an old 
suggestion that supermembranes
are string solitons or coherent states of interacting strings. It remains
to be seen if the strong coupling problem of string interactions
is tamed by the determination of the correct non-perturbative string vacuum.

\end{document}